\documentclass[
superscriptaddress,
 amsmath,amssymb,
 aps,
 prl,
 twocolumn,
floatfix,
longbibliography
]{revtex4-2}
\usepackage{graphicx}  
\usepackage{dcolumn}   
\usepackage{bm}        
\usepackage{amssymb}   
\usepackage{hyperref}
\usepackage{color}
\usepackage[utf8]{inputenc}

\begin{document}

\title{Electric Dipole Polarizability of $^{40}$Ca}

\newcommand{\TUDarmstadt}{Institut f\"ur Kernphysik, Technische Universit\"at Darmstadt, 64289 Darmstadt, Germany}
\newcommand{\OakRidge}{Physics Division, Oak Ridge National Laboratory, Oak Ridge, Tennessee 37831, USA}
\newcommand{\UTennessee}{Department of Physics and Astronomy, University of Tennessee, Knoxville, Tennessee 37996, USA}
\newcommand{\RCNP}{Research Center for Nuclear Physics, Osaka University, Ibaraki, Osaka 567-0047, Japan}
\newcommand{\FHU}{Faculty of Radiological Technology, Fujita Health University, Aichi 470-1192, Japan}
\newcommand{\EMMI}{ExtreMe Matter Institute EMMI, GSI Helmholtzzentrum f\"ur Schwerionenforschung GmbH, 64291 Darmstadt, Germany}
\newcommand{\MPIK}{Max-Planck-Institut f\"ur Kernphysik, Saupfercheckweg 1, 69117 Heidelberg, Germany}
\newcommand{\UM}{Institut f\"ur Kernphysik and PRISMA+ Cluster of Excellence, Johannes Gutenberg-Universität Mainz, 55128 Mainz, Germany}
\newcommand{\HIM}{Helmholtz-Institut Mainz, Johannes Gutenberg Universit\"at Mainz, 55099 Mainz, Germany}
\newcommand{\UCT}{Department of Physics, University of Cape Town, Rondebosch 7700, South Africa}
\newcommand{\Erlangen}{Institut f\"ur Theoretische Physik II, Universit\"at Erlangen, D-91058 Erlangen, Germany}
\newcommand{\Milano}{Dipartimento di Fisica, Universita` degli Studi di Milano, 20133 Milano, Italy and INFN, Sezione di Milano, 20133 Milano, Italy}
\newcommand{\MSU}{Facility for Rare Isotope Beams and Department of Physics and Astronomy, Michigan State University, East Lansing, Michigan 48824, USA}

\author{R.~W.~Fearick}\affiliation{\UCT}
\author{P.~von~Neumann-Cosel}\email[Email:]{vnc@ikp.tu-darmstadt.de}\affiliation{\TUDarmstadt}
\author{S.~Bacca}\affiliation{\UM}\affiliation{\HIM}
\author{J.~Birkhan}\affiliation{\TUDarmstadt}
\author{F.~Bonaiti}\affiliation{\UM}
\author{I.~Brandherm}\affiliation{\TUDarmstadt}
\author{G.~Hagen}\affiliation{\OakRidge}\affiliation{\UTennessee}
\author{H.~Matsubara}\affiliation{\RCNP}\affiliation{\FHU}
\author{W.~Nazarewicz}\affiliation{\MSU}
\author{N.~Pietralla}\affiliation{\TUDarmstadt}
\author{V.~Yu.~Ponomarev}\affiliation{\TUDarmstadt}
\author{P.-G.~Reinhard}\affiliation{\Erlangen}
\author{X.~Roca-Maza}\affiliation{\Milano}
\author{A.~Richter}\affiliation{\TUDarmstadt}
\author{A.~Schwenk}
\affiliation{\TUDarmstadt}\affiliation{\EMMI}\affiliation{\MPIK}
\author{J.~Simonis}\affiliation{\UM}
\author{A.~Tamii}\affiliation{\RCNP}


\begin{abstract}

The electric dipole strength distribution in $^{40}$Ca between 5 and 25~MeV has been determined at RCNP, Osaka, from proton inelastic scattering experiments at very forward angles. 
Combined with total photoabsorption data at higher excitation energy, this enables an extraction of the electric dipole polarizability $\alpha_\mathrm{D}(^{40}{\rm Ca}) = 1.92(17)$~fm$^3$.
Together with the measured $\alpha_{\rm D}$ in $^{48}$Ca, it provides a stringent test of modern theoretical approaches, including coupled cluster calculations with chiral effective field theory interactions
and state-of-the art energy density functionals. The emerging picture is that for this medium-mass region dipole polarizabilities are well described theoretically, with important constraints for the neutron skin in $^{48}$Ca and related equation of state quantities.
\end{abstract}


\maketitle 

{\em Introduction}.--
The nuclear equation of state (EOS) determines not only basic properties of nuclei~\cite{roc18} but also plays a key role for the properties of neutron stars and the dynamics of core-collapse supernovae and neutron star mergers~\cite{Lattimer21}. New observations from neutron stars and mergers provides constraints for the EOS of neutron-rich matter that can be compared with those derived from nuclear physics (see, e.g., Refs.~\cite{Raaijmakers21,ess21a,hut22}). 
However, while the EOS of symmetric nuclear matter is well determined around saturation density, the properties of neutron-rich matter are less explored experimentally.
The latter depends on the symmetry energy, whose properties are typically encoded in an expansion around saturation density $n_0$, with the symmetry energy at saturation density $J(n_0)$ and its density dependence  $L = 3 n_0 \partial J(n_0)/ \partial n$.

Theoretically, a model-dependent correlation between $L$ and the neutron-skin thickness $r_{\rm skin}$ in nuclei with neutron excess has been established~\cite{brown2000,cen09,rei10,roca-maza2011}. This correlation was also recently confirmed in {\it ab initio} computations of the neutron skin in $^{208}$Pb~\cite{hu22}.
Experimental attempts to determine the neutron skin thickness have been performed with a variety of probes (see, e.g., Ref.~\cite{thi19} and references therein), but many of them suffer from systematic uncertainties entering in the description of the reaction processes.
Parity-violating elastic electron scattering (a weak process mediated by the $Z^0$ boson) can be used for a nearly model-independent extraction of the neutron distribution in nuclei and, by comparison with accurately measured charge radii, the neutron skin thickness.
Recently, results with this technique have been reported by the CREX and PREX collaborations for $^{48}$Ca~\cite{adh22} and $^{208}$Pb~\cite{adh21}, respectively. 
The $r_{\rm skin}$ values inferred with selected nuclear models favor a comparatively small neutron skin in the former and a large skin in the latter case.

Alternatively, the electric dipole polarizability $\alpha_{\rm D}$ has been established as a possible measure of the neutron skin, based on the strong correlation with $r_{\rm skin}$~\cite{rei10,roc13}.
Data for $\alpha_{\rm D}$ extracted from proton inelastic scattering experiments at extreme forward angles have been presented for both
$^{48}$Ca~\cite{bir17}  and $^{208}$Pb~\cite{tam11}.
In these papers, two theoretical approaches have been used to describe $\alpha_{\rm D}$:
{\it ab initio} coupled-cluster (CC) calculations~\cite{Hagen2014,hag16} starting from chiral two- and three-nucleon interactions~\cite{Hebeler11,Ekstroem15} and energy density functional (EDF) theory~\cite{Bender2003}.

Attempts to simultaneously describe $\alpha_{\rm D}$($^{208}$Pb) and the parity-violating asymmetry from PREX and CREX with EDF models have shown limited success~\cite{rei21,pie21,rein2022,yuk22}. 
The values derived for $r_{\rm skin}$~\cite{adh21} and $L$~\cite{ree21} from PREX are in tension with EDFs capable of describing~\cite{roc15} the presently available results on $\alpha_{\rm D}$ in $^{48}$Ca~\cite{bir17}, $^{68}$Ni~\cite{ros13}, $^{120}$Sn~\cite{has15,bas20a}, and $^{208}$Pb~\cite{tam11}. While the CREX results is in excellent agreement with {\it ab initio} predictions~\cite{hag16}, the PREX result is in mild tension with the recent {\it ab initio} computations of $^{208}$Pb~\cite{hu22}.

Correlations between experimental observables and symmetry energy properties are well explored in EDF theory~\cite{brown2000,cen09,rei10,roc13}, but predictions for isovector observables like $\alpha_{\rm D}$ are less well constrained.
On the other hand, {\it ab initio} calculations provide a direct link to the EOS, as nuclear matter properties can be calculated based on the same chiral interactions~\cite{Hebeler11,hag16,dri19,hu22,Keller22}.
Results presented here are based on the set of two- and three-nucleon interactions from Refs.~\cite{Hebeler11,Ekstroem15} applied to study $\alpha_{\rm D}$ in $^{48}$Ca~\cite{hag16,bir17}.
The calculations of the $E1$ response are based on merging the Lorentz Integral Transform approach with CC theory, as described in Refs.~\cite{Bacca13,sim19}.
Recent work has extended the original two-particle--two-hole (2p-2h) CC truncation to include correlations up to three-particle--three-hole (3p-3h), so-called triples corrections, in the computation of $\alpha_{\rm D}$~\cite{mio18}. Their inclusion leads to a reduction of the predictions for $\alpha_{\rm D}$($^{48}$Ca) of the order of 10\%, allowing an improved simultaneous description of the charge radius~\cite{mio18}. A similar improvement was achieved for $^{68}$Ni~\cite{kau20}.

In this Letter, we present the measurement of the dipole polarizability
for $^{40}$Ca and confront it with CC and EDF calculations. This tests the emerging picture that nuclear theory can describe very well the neutron skin in medium-mass nuclei and related observables.

{\em Experiment}.-- 
Cross sections for the  $^{40}$Ca$(p,p')$ reaction have been measured at RCNP, Osaka, at an incident proton energy of 295\,MeV.  
Data were taken with the Grand Raiden spectrometer~\cite{fuj99} in a laboratory scattering angle range $0.4^\circ - 14.0^\circ$ and for excitation energies in the range $5 - 25$ MeV.
Dispersion matching techniques were applied to achieve an energy resolution of about 30~keV (full width at half maximum).  
The experimental techniques and the raw data analysis are described in Ref.~\cite{tam09}.

\begin{figure}
\includegraphics[width=\columnwidth,clip=]{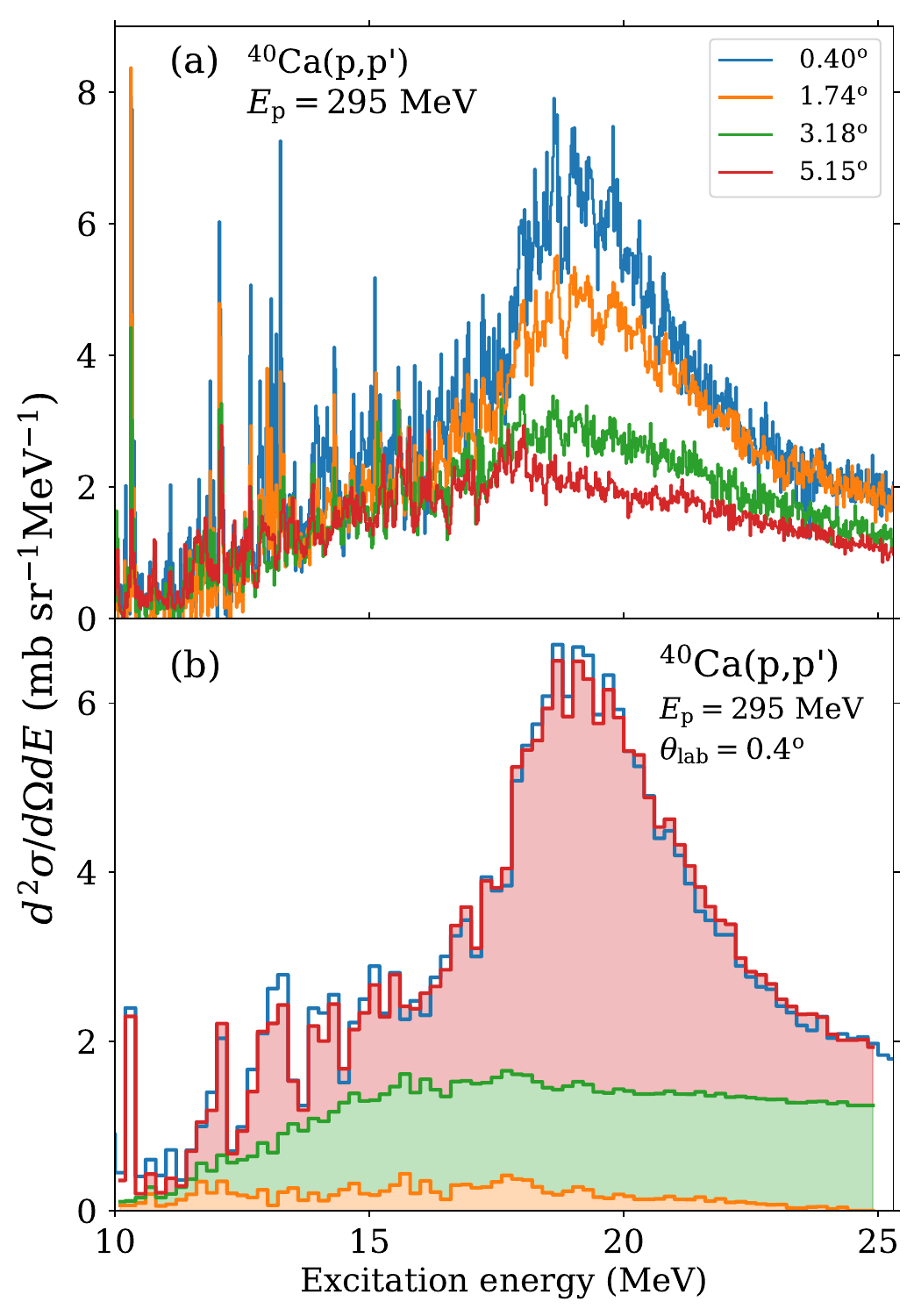}
\caption{
Top panel (a): Spectra of the $^{40}$Ca$(p,p')$ reaction at $E_0 = 295$~MeV and scattering angles $\Theta_{\rm lab} = 0.4^\circ$, $1.74^\circ$, $3.18^\circ$ and $5.15^\circ$.
Bottom panel (b): Example of the MDA of the spectrum at $\Theta_{\rm lab} = 0.4^\circ$ in 200 keV bins (blue) and decomposition into contributions of $\lambda \neq 1$ multipoles (orange), continuum background (green), and $E1$ (red). 
\label{fig:spectra}}
\end{figure}

In the top panel of Fig.~\ref{fig:spectra} we show representative energy spectra measured at laboratory scattering angles $\Theta_{\rm lab} = 0.4^\circ$, $1.74^\circ$, $3.18^\circ$, and $5.15^\circ$. The predominant cross sections lie in the energy region above 10 MeV.
$M1$ strength in $^{40}$Ca is known to be concentrated in a single prominent transition at 10.32 MeV \cite{gro79}.
The cross sections above 10 MeV show a broad resonance structure peaking at about 19 MeV increasing towards  $0^\circ$. The angular dependence  is consistent with relativistic Coulomb excitation of $E1$ transitions.
We identify this resonance structure as the isovector giant dipole resonance.

The various contributions to the spectra were separated using a multipole decomposition analysis (MDA) as described in Ref.~\cite{vnc19}.
Results for the most forward angle measured are presented in the bottom part of Fig.~\ref{fig:spectra} as example, where the spectra was rebinned to 200 keV.
Theoretical angular distributions for the relevant multipoles were obtained from Distorted Wave Born Approximation calculations with transition amplitudes from quasiparticle-phonon-model calculations similar to the analysis of $^{48}$Ca \cite{bir17}. 
Additionally, a background due to pre-equilibrium multistep scattering was considered. 
Its angular dependence was taken from experimental systematics \cite{kalbach81,chad94} while the amplitude was derived by two means. 
Initially, an unconstrained fit was done at each energy bin of the set of spectra. 
The resulting cross sections could be well approximated by a simple Fermi function but showed strong fluctuations for certain excitation energy bins due to the similarity to some of the $E1$ theoretical angular distributions. 
Thus, in the final analysis, the continuum contribution was determined by fitting a Fermi function to the unconstrained excitation energy dependence.

\begin{figure}
\includegraphics[width=\columnwidth,clip=]{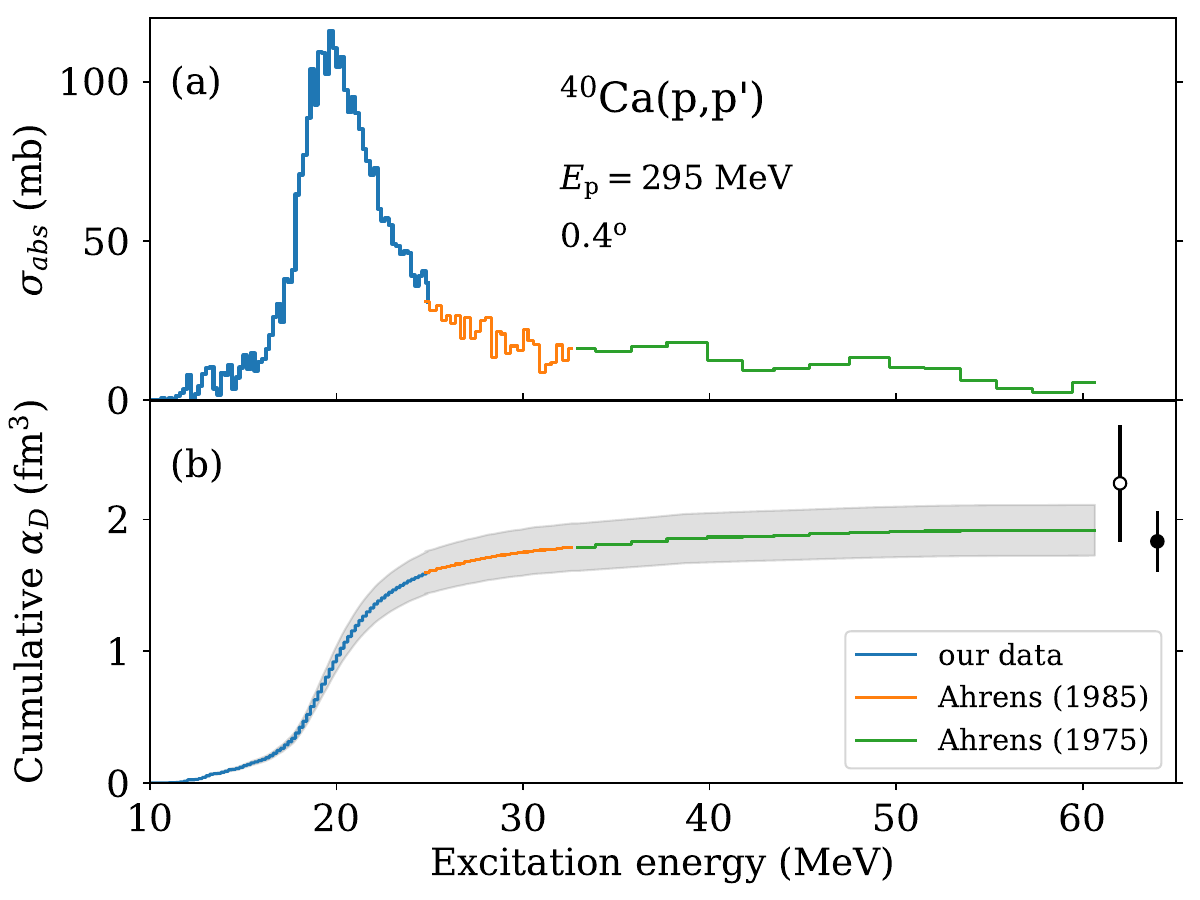}
\caption{
Top panel (a): Photoabsorption cross section derived at a scattering angle of $0.40^\circ$ using the virtual photon method.
Bottom panel (b): Electric dipole polarizability $\alpha_D$ derived from the photoabsorption cross sections. The blue curve shows the present data, while the orange and green curves show the extrapolation to higher energies using the data of Refs.~\cite{ahr75,ahr85}.
The open (full) black circles are the CC results for the NNLO$_{\rm sat}$ interaction including up to doubles (triples) contributions in the cluster expansion.
\label{fig:absalpha}}
\end{figure}

{\em Photoabsorption cross sections and dipole polarizability}.--
The $E1$ cross sections resulting from the MDA were converted into equivalent photoabsorption cross sections using the virtual photon method~\cite{ber88}.  
The virtual photon spectrum was calculated in an eikonal approach \cite{ber93} to Coulomb excitation, integrated over the distribution of scattering angles covered in the solid angle of each angular bin. 
The photoabsorption spectra derived from scattering data  at $0.40^\circ$ and $1.00^\circ$ were essentially identical, and that at $1.74^\circ$ deviated only slightly, consistent with an estimate of the grazing angle ($1.33^\circ$) at which Coulomb-nuclear interference becomes relevant.
The resulting photoabsorption cross section is displayed as blue histogram in Fig.~\ref{fig:absalpha}~(a).

The electric dipole polarizability $\alpha_D$ was obtained from the photoabsorption cross section over the energy range $10-25$ MeV leading to a contribution 1.60(14)~fm$^3$. The integration was extended to 60 MeV, where the cumulative sum plotted in Fig.~\ref{fig:absalpha}~(b) shows saturation. The data at higher excitation energies were taken for $25-31$ MeV from Ref.~\cite{ahr85} and for $31-60$ MeV from Ref.~\cite{ahr75} to obtain the total $\alpha_D(^{40}{\rm Ca}) = 1.92(17)$~fm$^3$.
The uncertainty considers systematic errors of (i) the absolute cross sections,  (ii) the MDA (determined as described, e.g., in Ref.~\cite{bassauer20a}), and (iii) the parameterization of the continuum background (evaluated by varying the amplitude of the Fermi function), added in quadrature. The latter, dominating the total uncertainty budget, was estimated by the variation needed to change the $\chi^2$ value of the MDA fit by one. Statistical errors turned out to be negligible. A detailed breakdown of the error contributions is given in Table \ref{tab:errors}.

\begin{table}
\caption{Budget of error contributions to $\alpha_{\rm D}$($^{40}$Ca).}	
\label{tab:errors}
\begin{ruledtabular}
\begin{tabular}{lc}
Source & Value (\%) \\
\hline
Trigger efficiency & 0.1 \\
Drift chamber efficiency & 0.8 \\
Charge collection & 0.3 \\
Target thickness & 1.0 \\
Determination of solid angle & 3.0 \\
MDA  & 1.2 \\
Background parameterization & 8.3 \\
\hline
Total & 9.0 \\
\end{tabular}
\end{ruledtabular}
\end{table} 

{\em Comparison with coupled-cluster calculations}.--
The extracted value of $\alpha_{\rm D}$ serves as a benchmark for CC theory~\cite{Bacca13,hag16,mio18,sim19}. 
Coupled-cluster calculations were recently performed for the dipole polarizability of $^{48}$Ca~\cite{bir17} and $^{68}$Ni~\cite{kau20}, which led to an improved understanding of the neutron and proton distributions in nuclei, as well as their difference encoded in the neutron skin.  We have performed CC computations of $\alpha_{\rm D}$ in $^{40}$Ca starting from a Hartree-Fock reference state considering a basis of 15 major harmonic oscillator shells. To gauge the convergence of our results we varied the oscillator frequency in the range $\hbar\omega = 12-16$ MeV. Three-nucleon contributions had an additional energy cut of $E_{\rm 3max} = 16 \hbar\omega$.

\begin{figure}[t]
\includegraphics[width=\columnwidth,clip=]{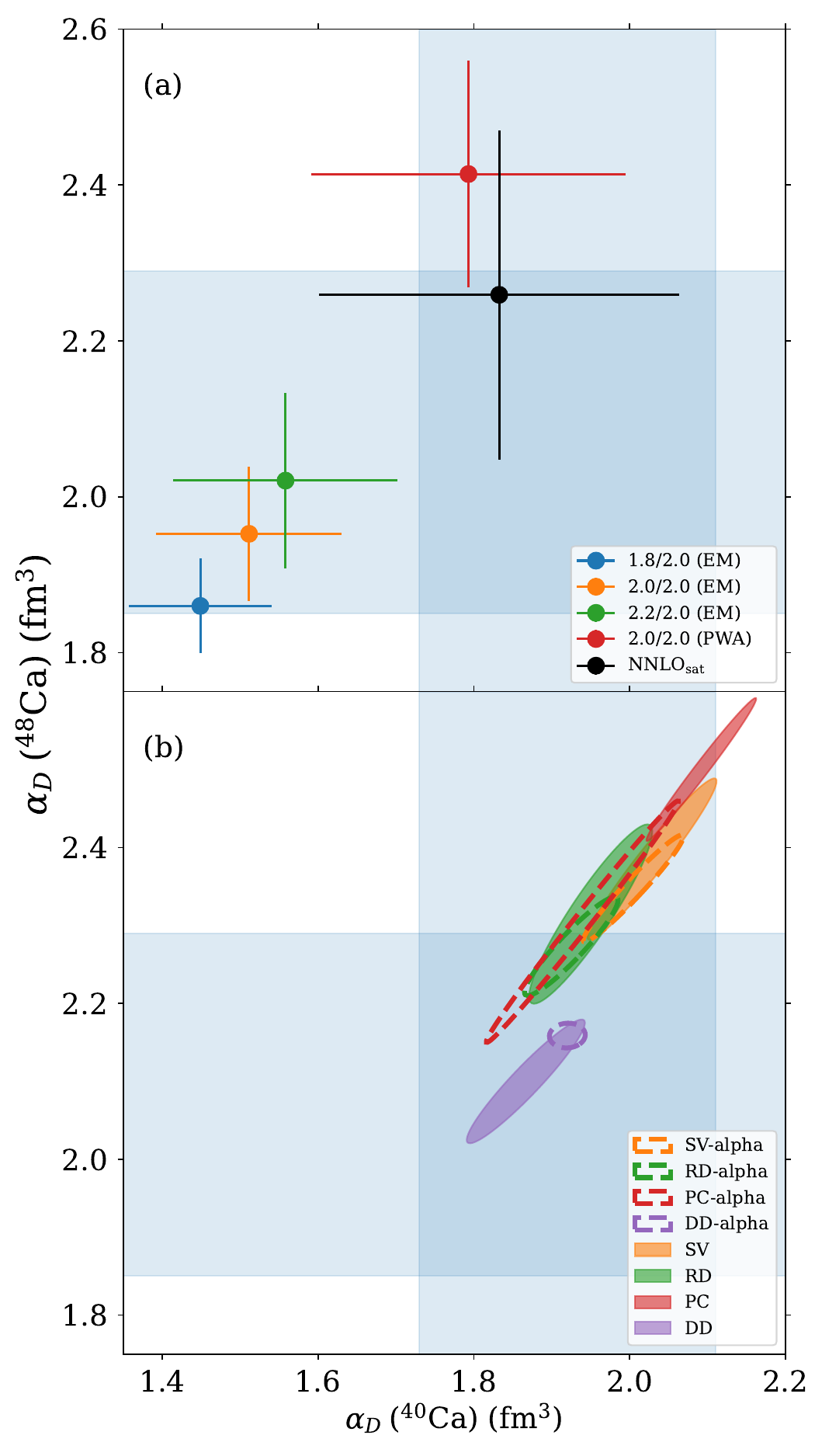}
\caption{
Comparison of the experimental dipole polarizabilites of $^{40}$Ca (present work) and $^{48}$Ca \cite{bir17} shown as blue bands with (top panel, a) CC calculations with different interactions, including triples contributions and (bottom panel, b) EDF calculations  with different energy density functionals~\cite{rei21}. For details see text.
\label{fig:40Cavs48Ca}
}
\end{figure}

Figure~\ref{fig:40Cavs48Ca} explores the correlation between $\alpha_{\rm D}$ for $^{40}$Ca and $^{48}$Ca as predicted by theory. Panel (a) shows the CC results including triples  contributions, not available for  $^{40}$Ca so far. The theoretical uncertainties for the different Hamiltonians stem from the truncation of the CC expansion and the residual dependence on CC convergence parameters, calculated as described in Ref.~\cite{sim19}. 
Similarly to $^{48}$Ca, we find that the inclusion of 3p-3h correlations reduces the value of $\alpha_{\rm D}$($^{40}$Ca) by an amount varying between 10\% -- 20\% for  different interactions. While the EM and PWA interactions~\cite{Hebeler11} are not simultaneously compatible with both $^{40}$Ca and $^{48}$Ca experimental data, the set of employed interactions shows an approximately linear trend  between the two quantities overlapping with both experimental results.
A particular improvement in the reproduction of both $\alpha_{\rm D}$($^{48}$Ca) and $\alpha_{\rm D}$($^{40}$Ca) is seen for the NNLO$_{\rm sat}$ interaction~\cite{Ekstroem15}, which is capable of accurately describing binding energies and radii of nuclei up to $^{40}$Ca as well the saturation point of symmetric nuclear matter.
The different interactions predict a range of symmetry energy parameters $J = 27 - 33$~MeV, $L = 41 - 49$ MeV~\cite{hag16}, with the NNLO$_{\rm sat}$ values at the lower end ($J = 27$~MeV, $L = 41$~MeV).

{\em Comparison with EDF approaches}.--
Recently, it was investigated whether the dipole polarizability  and the parity-violating asymmetry $A_{\rm PV}$  for $^{208}$Pb and $^{48}$Ca can be simultaneously accounted for with modern EDFs \cite{rein2022}.
We use the four representative forms of functionals from that study: non-relativistic Skyrme functionals SV \cite{klu09} and RD \cite{erl10}, the latter with different forms of density dependence, and relativistic functionals DD \cite{nik02} with finite-range meson-exchange coupling and PC \cite{nik08} with point coupling. All four have been calibrated to the same set of ground-state data to determine the model parameters. With these sets, it was shown that PREX and CREX results for $A_{\rm PV}$ (and $r_{\rm skin}$) cannot be consistently explained  within the model uncertainties while the $\alpha_\mathrm{D}$ were reproduced. 
Hence, the present result in $^{40}$Ca provides an important test of the global predictive power of these EDFs.

Figure~\ref{fig:40Cavs48Ca}~(b) displays the EDF results for $\alpha_{\rm D}$ with $1\sigma$ error ellipses (for their definition see Refs.~\cite{rei21,rein2022}). The parametrizations as given from the ground-state fits are shown by filled ellipses.
The DD functional performs rather well. The other predictions tend to  slightly overestimate the experimental mean values of both $^{40}$Ca and $^{48}$Ca, while their $1\sigma$ error ellipses do overlap with the experimental bands, except for PC. In all cases, the two $\alpha_\mathrm{D}$ values are highly correlated. We note that the same holds for the description of $\alpha_D$($^{208}$Pb) \cite{tam11} after correction for the quasi-deuteron contribution \cite{roc15}. Thus, all the models are capable to account for the mass dependence of the polarizability.

The dashed ellipses show results from a refit where additionally the experimental $\alpha_\mathrm{D}$ value of $^{208}$Pb \cite{tam11} corrected for the quasideuteron part \cite{roc15} was included yielding the functionals SV-alpha; RD-alpha, PC-alpha, and DD-alpha \cite{rei21,rein2022}. This improves the agreement with experiment, particularly for the PC model, and shrinks most error ellipsoids. The uncertainty reduction is especially large for the DD model because this functional has the least isovector freedom. The linear trend shown by the different theoretical approaches in Fig.~\ref{fig:40Cavs48Ca} is similar although the CC calculations tend to underestimate the $\alpha_D$ in $^{40}$Ca and perform nicely for $^{48}$Ca. The bulk symmetry energies range from $J=30$\,MeV for DD to 35\,MeV for PC and accordingly from 
32\,MeV to 82\,MeV for $L$. The fits which include also $\alpha_\mathrm{D}$ in $^{208}$Pb narrow the prediction to $J=30-32$ MeV and $L=35-52$\,MeV which correlates nicely to the narrower range of predictions for $\alpha_\mathrm{D}$ in $^{40,48}$Ca.

{\em Conclusions}.--
We have extracted the dipole polarizability of $^{40}$Ca from a combination of relativistic Coulomb excitation measurement in inelastic proton scattering under very forward angles with total photoabsorption data at high excitation energies.
Together with a similar analysis on $^{48}$Ca the new data serve as a benchmark test of state-of-the art theoretical approaches.
A representative set of EDFs can describe these data. An improvement is obtained  when the EDFs are optimized  by adding the dipole polarizability of $^{208}$Pb to the calibration dataset.
Coupled-cluster computations for the NNLO$_{\rm sat}$ interaction simultaneously describe well the dipole polarizability of $^{40}$Ca and $^{48}$Ca, as well as the corresponding charge radii and the neutron skin thickness~\cite{sim19}. A nearly linear systematic trend is obtained for other interactions, as in the case of EDF theory.
This analysis supports the robustness of current theoretical approaches in the description of $\alpha_D$ and their constraints of symmetry energy parameters discussed, e.g., in Refs.~\cite{rei21,rein2022,lat23}. 


\begin{acknowledgments}
This work was supported by the Deutsche Forschungsgemeinschaft (DFG, German Research Foundation) -- Project-ID 279384907 –- SFB 1245, through the Cluster of Excellence ``Precision Physics, Fundamental Interactions, and Structure of Matter" (PRISMA$^+$ EXC 2118/1, Project ID 39083149), by the U.S.\ Department of Energy, Office of Science, Office of Nuclear Physics under award numbers DE-SC0013365 and DE-SC0023175 (NUCLEI SciDAC-5 collaboration), under the contract DE-AC05-00OR22725 with UT-Battelle, LLC (Oak Ridge National Laboratory), and by the University of Cape Town. Computer time was provided by the Innovative and Novel Computational Impact on Theory and Experiment (INCITE) programme. This research used resources of the Oak Ridge Leadership Computing Facility at the Oak Ridge National Laboratory, which is supported by the Office of Science of the U.S.\ Department of Energy under Contract No. DE-AC05-00OR22725.
\end{acknowledgments}

\bibliography{references_40Ca}

\end{document}